\RequirePackage{fix-cm}
\documentclass[smallextended]{svjour3}       
\smartqed  
\usepackage{graphicx}
\usepackage{amsmath,amssymb,mathtools,booktabs}

\usepackage{units,dsfont}

\newcommand{\Ft}{\mathbf{F}}

\newcommand{\Id}{\mathbf{I}}

\newcommand{\ut}{\mathbf{u}}
\newcommand{\vt}{\mathbf{v}}
\newcommand{\nt}{\vec{n}}
\newcommand{\taut}{\boldsymbol{\tau}}

\newcommand{\ft}{\mathbf{f}}
\newcommand{\sigmat}{\boldsymbol{\sigma}}
\newcommand{\Sigmat}{\boldsymbol{\Sigma}}

\renewcommand*{\thesection}{\Roman{section}}

\newcommand{\ve}[1]{\mathbf{#1}}
\newcommand{\te}[1]{\mathbf{#1}}

\newcommand{\C}{\te{C}}

\newcommand{\n}{\ve{n}}

\def\wss{\text{WSS}}

\def\rho{\varrho}

\newcommand{\tr}{\operatorname{tr}\,}

\renewcommand{\div}{\operatorname{div}\,}

\newcommand{\SO}{\mathcal{S}}
\newcommand{\IN}{\mathcal{I}}

\newcommand{\FL}{\mathcal{F}}
\newcommand{\xt}{\mathbf{x}}

\newcommand{\phit}{\boldsymbol{\phi}}

\renewcommand{\div}{\operatorname{div}}

\usepackage[usenames,dvipsnames]{xcolor}
\usepackage{tikz-cd}
\usetikzlibrary{calc,arrows,positioning}
\usepackage{pgfplots}
\pgfplotsset{compat=1.14}

\begin{document}

\title{On the Impact of Fluid Structure Interaction \\in Blood Flow Simulations 
}
\subtitle{Stenotic Coronary Artery Benchmark}


\author{Lukas Failer         \and Piotr Minakowski \and Thomas Richter   }


\institute{L. Failer \at
	Siemens AG, Corporate Technology, 80200 Munich, Germany \\
              \email{lukas.failer@siemens.com}           
           \and
           P. Minakowski \at
Institute of Analysis and Numerics, Otto-von-Guericke
University Magdeburg, Universit\"atsplatz 2, 39106 Magdeburg,
Germany,\\
\email{  piotr.minakowski@ovgu.de}
\and
T. Richter \at
Institute of Analysis and Numerics, Otto-von-Guericke
University Magdeburg, Universit\"atsplatz 2, 39106 Magdeburg,
Germany, 
and Interdisciplinary Center for Scientific Computing, 
Heidelberg University, INF 205, 69120 Heidelberg, Germany, \\
\email{  thomas.richter@ovgu.de}
}

\date{Received: date / Accepted: date}

\maketitle

\begin{abstract}
    
We study the impact of using fluid-structure interactions (FSI)
to simulate blood flow in a large stenosed artery.
We compare typical flow configurations using Navier-Stokes
in a rigid geometry setting to a fully coupled FSI model.
The relevance of vascular elasticity is investigated with
respect to several questions of clinical importance. 
Namely, we study the effect of using FSI on the wall shear
stress distribution, on the Fractional Flow Reserve and on the damping
effect of a stenosis on the pressure amplitude during the pulsatory
cycle. 
The coupled problem is described in a monolithic variational
formulation based on Arbitrary 
Lagrangian Eulerian (ALE) coordinates. For comparison, we perform pure
Navier-Stokes simulations on a prestressed geometry to give a good
matching of both configurations. A series of numerical 
simulations that cover important hemodynamical factors are presented and discussed.

\end{abstract}

\section{Introduction}\label{sec:intro}

The application of computational fluid dynamics (CFD) to blood flow is a rapidly
growing field of biomedical and mathematical research. Currently the development 
of numerical methods in hemodynamics is evolving from a purely academic tool
to aiding in clinical decision making \cite{Taylor1999,Bluestein2017}.
In particular the investigation of blood flow in stenosed arteries can help to shape medical
treatment. For example virtual/computed Fractional Flow Reserve (cFFR), can evaluate the
physiological significance of sclerotic plaque \cite{Morris2015,Blanco2018}. 
Moreover, the correct reconstruction of wall shear stress (WSS) 
is of crucial importance for the cell signaling and in
consequence for the stenosis development \cite{Liang2009} or for the assessment 
of rupture of brain aneurysms\cite{Cebral2010}. These two factors are studied in detail. 

Hemodynamical simulations face numerous challenges, that are connected 
with measurements, medical image segmentation, mathematical modelling 
and development of numerical methods ~\cite{Quarteroni2000}.
In this work we confine to an idealized geometry and 
a Newtonian fluid and focus on the comparison between
a compliant wall vs. a rigid vessel wall. 
As a model example we have chosen a curved channel that resembles a large human 
artery. Stenotic and non-stenotic configurations are considered.
The channel is preloaded by first considering a steady inflow to reach
a certain physiological pressure and diameter 
before starting the pulsatile hear cycle. We investigate the
aforementioned important clinical hemodynamical factors cFFR and
WSS.

In the physiology of vessel macrocirculation the compliance plays a crucial role. 
However for individual arteries the problem is still open and there is no gold standard 
that can be adapted into clinical practice. For a discussion on assumptions 
in hemodynamic modeling  of large arteries we refer to \cite{Steinman2012}.

Numerous numerical studies were performed to compare rigid with compliant vessel simulations.  
The results report reduction of WSS for compliant vessels compared to rigid walls. 
In the case of a carotid bifurcation the authors of \cite{Perktold1995} could 
not observe significant changes in the flow patterns. In contrast relatively 
large vessels can undergo substantial radial wall motion, 
for example in the case of coronary arteries.
The motion consists of bulk deformation and wall compliance that 
results in notable changes of flow characteristics \cite{Jin2003,Torii2010}.

After the brief introduction in Section \ref{sec:intro}  we present
the monolithic formulation of the FSI problem, see Section \ref{sec:model}. 
The setting of the simulations is the topic of Section~\ref{sec:setting}. 
The numerical method is described in Section \ref{sec:nummethod}.
Section \ref{sec:sim} is dedicated to the presentation of investigated
hemodynamical factors and the discussion of  numerical results 
obtained on relevant test cases.
Finally in Section \ref{sec:con} we present a conclusion of this work.

\section{Model description}\label{sec:model}

We consider a $3$-dimensional domain $\Omega\subset\mathds{R}^3$, that
represents a part of a vessel.  The domain is partitioned in the
reference configuration 
$\Omega = \FL\cup\IN\cup\SO$, where $\FL$ is the fluid domain, $\SO$
the solid domain and $\IN=\partial\FL\cap\partial\SO$ is the fluid
structure interface.

The velocity field $\vt$ and the deformation field $\ut$ are split
into fluid  $\vt_f\coloneqq\vt|_{\FL}$, $\ut_f\coloneqq\ut|_{\FL}$ and solid
$\vt_s\coloneqq\vt|_{\SO}$,   $\ut_s\coloneqq\ut|_{\SO}$ counterparts
respectively. The pressure variable $p_f$ only exists on the fluid domain.  

The boundary of the fluid domain $\Gamma_f\coloneqq\partial\FL\setminus\IN$ is
split into the inflow boundary $\Gamma_f^{in}$ and the outflow
boundary~$\Gamma_f^{out}$. 
Similarly the solid boundary $\Gamma_s=\partial\SO\setminus\IN$ is split
into inflow $\Gamma_s^{in}$ and outflow $\Gamma_s^{out}$ boundaries.
Boundary conditions are described in Section~\ref{sec:setting}.

\subsection{Fluid material model}

Although, blood exhibits many unique properties, i.e. in certain regimes 
blood shows a non-Newtonian behaviour, we confine this work to considering
 an incompressible Newtonian fluid with the viscosity of 
$\mu_f = \unit[0.033]{g\cdot cm^{-1} s^{-1}}$ and the density $\rho_f= 
\unit[1]{g\cdot cm^{-3}}$. 
The assumption of a Newtonian model for blood rheology is widely 
accepted for large and medium vessels, see e.g. \cite{Quarteroni2000}. 
The flow is governed by the Navier-Stokes equations
\begin{subequations}\label{ns}
  \begin{align}
    \rho_f\left(\partial_t  \vt_f +  (\vt_f \cdot\nabla) \vt_f \right)
    -  \mu_f\div\big(\nabla\vt_f+\nabla\vt_f^T\big)  +
    \nabla p_f & = 0 \quad \text{in } \FL, \\  
    \div \vt_f&= 0\quad \text{in } \FL.
  \end{align}
\end{subequations}

\subsection{Solid material model}

Arterial walls consist of heterogeneous layers with significant difference in physical properties. 
The principle layer construction of arterial wall consist of intima (inner layer), media (middle layer), and  
adventitia (outer layer). For detail account we refer to~\cite{fung1993biomechanics} 
and~\cite{holzapfel2000new}. We briefly describe how the elastic constitutional law used in this work is derived.

Since arteries hardly change their volume within the physiological range of deformation \cite{Carew1968}, 
they can be regarded as incompressible or nearly incompressible materials. This motivates the application of 
a multiplicative decomposition of the deformation tensor ($\Ft$) into 
its volumetric  part $J^{\frac{1}{3}}$ and the deviatoric part $\bar{\Ft}$:
\begin{align}
\Ft = \Id+\nabla\ut_s,\quad \Ft=J^{\frac{1}{3}} \bar \Ft, \text{ where } J = \det \Ft.
\end{align}
Its associated modified deviatoric Cauchy Green tensor $\bar \C$ then has the structure 
\begin{align}
\bar \C=\bar \Ft^T\bar \Ft.
\end{align}

Thereby the free-energy function $\Psi$ can be split in a volumetric and 
deviatoric part as described in  \cite{holzapfel2000new} or \cite{Gurtin2010}: 

$$\Psi =\Psi_{VOL}(J)+\Psi_{DEV}(\bar \C).$$

Artery and vein walls consist of elastin and colagen fibres.
The measurements of stress-strain curve exhibit 
stiffening effects at higher pressures due to  the collagen 
fibres, c.f. \cite{fung1993biomechanics}. Whereas under low 
loading of the artery the properties of elastin dominates. 
This motivates to model the artery as pseudo-elastic material. 
Following \cite{Delfino1997} and \cite{holzapfel2000new} 
we employ an exponential  deviatoric energy functional 
\begin{align}
\Psi_{DEV}(\bar C)= \frac{\mu}{2\gamma} \Big(\exp^{\gamma (tr(\bar C)-3)}-1\Big).\label{eq_energy_deviatoric}
\end{align}
For the volumetric part of the energy functional the simple convex energy function 
\begin{align}
\Psi_{VOL}(J)&= \frac{\kappa}{2} \left(\frac{1}{2} (J^2-1)-\ln(J)\right)\label{eq_energy_vol_2}
\end{align}
is used as stated in \cite{bazilevs2010computational,bazilevs2008isogeometric}.

By assuming hyperelastic stress response we obtain the second Piola–Kirchhoff stress tensor $\Sigmat_s$: 
\begin{equation}\label{eq_stress_strain_quarteroni}
\begin{split}
\Sigmat_s(J,\Ft)=& \frac{\partial \Psi_{VOL}(J)}{\partial \C}
  +\frac{\partial \Psi_{DEV}(\bar \C)}{\partial \C} \\
 =& \mu_s J^{-2/3}(\Id-\frac{1}{3} \tr(\Ft^T \Ft) (\Ft^{T}\Ft)^{-1}) e^{\gamma (J^{-2/3} \tr(\Ft^T \Ft)-3)} \\
  &+ \frac{\kappa_s}{2}(J^2-1) J (\Ft^{T}\Ft)^{-1} ,
\end{split}
\end{equation}
with the material parameters $ \mu_s=\unit[44.2]{kPa}$ and $\gamma=20$
as well as $\kappa_s=\unit[4998]{kPa}$. 
Similar values have been used in \cite{Balzani2016}.
Moreover the solid density equals $\hat\rho_s= \unit[1.2]{g\cdot
  cm^{-3}}$.

The equation for the conservation of momentum in the
elastic vessel wall is then given by
\begin{subequations}\label{solid}
  \begin{align}
    \hat\rho_s d_{t}\vt_s - \div\Big(\Ft\Sigmat_s\Big) &=\hat\rho_s \ft_s \quad \text{in }\SO,\\
    d_t \ut_s&=\vt_s \quad \text{in }\SO,
  \end{align}
\end{subequations}
where we formulated the hyperbolic problem as a first order system in time by
introducing the solid velocity $\vt_s$ and the deformation field $\ut_s$ as separate
variables. The solid problem is formulated in the Lagrangian coordinates
on the reference domain $\SO$, which is not moving with time. Hence,
the density $\hat\rho_s=\unit[1.2]{g\cdot cm^{-3}}$ is the reference
density which is unaffected by any compression or extension.

\subsection{Fluid-structure interactions}

One aim of this work is to study the impact of coupled fluid-structure
interactions on typical blood flow configurations found in medical
applications.
We therefore couple the Navier-Stokes equations~(\ref{ns})
with the elastic material law~(\ref{solid}) via coupling conditions on
the common interface $\IN$. The coupled dynamics of a fluid-structure
interaction problem leads to a free boundary problem with moving
domains and a moving interface. 

In the following, we briefly sketch the derivation of the coupled
fluid-structure interaction problem. For details we refer to the
literature~\cite[Chapter 5]{Richter2017}. 
To overcome the mismatch of a 
Navier-Stokes equations~(\ref{ns}) defined in Eulerian coordinates, e.g. on
the evolving fluid domain $\FL(t)$, and the solid
problem~(\ref{solid}) derived on the fixed reference domain $\SO$ we use the well established concept of
Arbitrary Lagrangien Eulerian (ALE) coordinates. See ~\cite{Donea1982}
or~\cite[Chapter 5]{Richter2017} for a detailed derivation. We denote
by $\FL$ the fixed 
fluid reference domain and by $T_f(t):\FL\to\FL(t)$ the ALE map. Then,
the velocity and the pressure can be mapped onto the fixed domain by defining
$\hat\vt_f\coloneqq\vt_f\circ T_f^{-1}$ and $\hat p_f\coloneqq
p_f\circ T_f^{-1}$.
This allows to transform the Navier-Stokes problem
in its variational formulation onto the fluid reference domain
\begin{equation}\label{fluid-ale}
\begin{split}
\Big(J_f\big(\partial_t  \hat\vt_f+(\Ft_f^{-1}(\vt-\partial_tT_f)
\cdot\hat\nabla) \hat\vt_f\big),\phit\Big)_\FL+
\Big(J_f\hat\sigmat_f\Ft_f^{-T},\hat\nabla\phit\Big)_\FL\\
+\Big(J_f\Ft_f^{-1}:\hat\nabla\hat\vt_f^T,\xi\Big)_\FL =
\Big(J_f\rho_f \hat\ft,\phit\Big)_\FL,
\end{split}
\end{equation}
where $\phit$ and $\xi$ are test functions.  We denote by $\Ft_f\coloneqq \hat\nabla T_f$ the
gradient of the deformation variable and by $J_f\coloneqq \det(\Ft_f)$ its
determinant. Most characteristic
feature of the ALE formulation is the appearance of the domain
convection term $-(\Ft_f^{-1}\partial_t T_f\cdot\nabla)\hat\vt$ that
takes care of the implicit motion of the fluid domain. Furthermore, the
Cauchy stress tensor is mapped to the reference domain which gives
rise to the Piola transform $J_f\hat\sigmat_f\Ft_f^{-T}$. The
reference stress is given by
\[
\hat\sigmat_f(\vt,p) = -\hat p_f \Id + \rho_f\nu_f (\hat\nabla\hat\vt\Ft^{-1} + \Ft^{-T}\hat\nabla\hat\vt^T).
\]

It remains to describe the construction of the ALE map. 
Typically the ALE map is defined by means of an artificial fluid domain deformation
$\ut_f$ via
\[
T_f(\xt,t)\coloneqq\xt+\hat\ut_f(\xt,t),
\]
where $\hat\ut_f$ is an extension of the solid deformation $\ut_s$
from $\SO$ to the fluid reference domain $\FL$. The most simple choice
for the extension operator is to use a harmonic extension by implicitly solving the vector
Laplacian
\begin{equation}\label{extension}
  -\hat\Delta\ut_f = 0\text{ in }\FL,\quad
  \ut_f=\ut_s\text{ on }\IN,\quad
  \ut_f=0\text{ on }\partial\FL\setminus \IN. 
\end{equation}
For a discussion of this extension operator we refer to the
literature~\cite[Sections 3.5.1 and 5.3.5]{Richter2017}. Hereby, $T_f$
can be considered as a natural extension of the Lagrange-Euler map
$T_s(\xt,t)\coloneqq \xt+\ut_s(\xt,t)$ such that we will skip the
subscripts $f$ and $s$ when denoting the deformation
$T(\xt,t)\coloneqq \xt+\ut(\xt,t)$, its 
gradient $\Ft=\nabla T$ and its determinant $J=\det(\Ft)$. 
In the following we skip all hats referring to the use of ALE
coordinates.

As the fluid reference domain $\FL$ and the Lagrangian solid domain
$\SO$ do not move, they always share the well defined common interface
$\IN$. Here, we require the continuity of velocities, which is denoted
by the \emph{kinematic coupling condition}
\[
\vt_f=\vt_s\text{ on }\IN,
\]
continuity of normal stresses, denoted as \emph{dynamic coupling
  condition}
\[
\Ft\Sigmat_s\nt=J\sigmat_f\Ft^{-T}\nt\text{ on }\IN,
\]
and finally, the \emph{geometric coupling condition} which says that
the evolving domains $\FL(t)$ and $\SO(t)$ may not overlap and may not
separate at the interface. 

Since the velocity field and the deformation field are continuous across the
interface, we formulate the coupled fluid-structure interaction
problem using global solution fields $\vt\in H^1(\Omega)^3$ and
$\ut\in H^1(\Omega)^3$. Hereby, the kinematic coupling condition and
the extension condition $\ut_f=\ut_s$ in~(\ref{extension}) are strongly
realized as parts of the function spaces. The dynamic coupling
condition is realized by testing the variational formulations of the 
Navier-Stokes equations an the solid problem by one common continuous
test function $\phit\in H^1(\Omega)^3$. The resulting variational system of
equations is given by
\begin{subequations}\label{probelm}
  \begin{align}
    \begin{split}
      \big( J(\partial_t  \vt +  (\Ft^{-1}( \vt-\partial_t   
      \ut)\cdot\nabla) \vt,\phit\big)_{\FL} +& \big( J\hat\sigmat_f
      \Ft^{-T},\nabla\phit\big)_{\FL}\hspace{-1cm} \\
      +(\hat\rho_s\partial_t \vt,\phit)_{\SO}
      +( \Ft\Sigmat_s,\nabla\phit)_{\SO} &=
      ( J\rho_f \ft,\phit)_{\FL} 
      +(\hat\rho_s \ft,\phit)_{\SO}
    \end{split},\\
    \big(J\Ft^{-1}:\nabla\vt^T,\xi\big)_{\FL} &= 0,\\
    (\partial_t \ut- \vt,\psi_s)_{\SO}&=0,\\
    (\nabla \ut,\nabla\psi_f)_{\FL}&=0.
  \end{align}
\end{subequations}
For detailed account of monolithic formulations for fluid structure
interactions we refer to~\cite{Richter2017}.

\section{Simulation setup}\label{sec:setting}

We perform pulsatile blood-flow simulations by numerically solving 
system~\eqref{probelm}. The simulation setup that includes geometry and material parameters 
has been inspired by the Benchmark paper~\cite{Balzani2016}. Note however that in some of the cases the assumption of rigid vessel walls is employed
such that $\ut = 0$ and $\Ft=\Id$ on $\Omega$. Then system \eqref{probelm}
reduces to the standard incompressible Navier-Stokes equations. 
In what follows we distinguish between FSI- and NS-cases,
respectively.



\begin{figure}[t]
    \centerline{\includegraphics[width=0.6\textwidth]{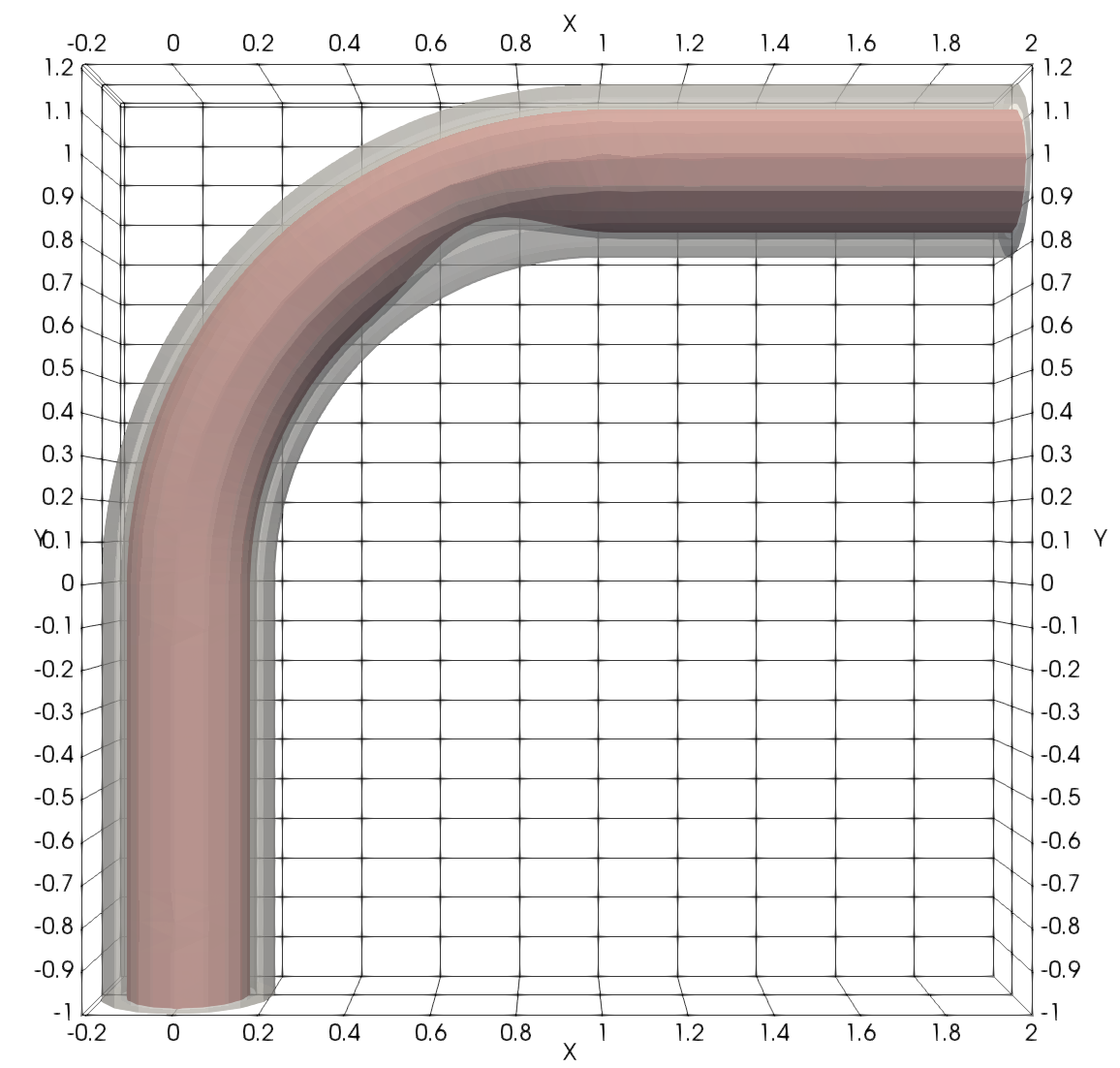} }
    \caption{Computational domain showing Geometry 3 with a
      non-symmetric stenosis. }
    \label{fig:domain}
\end{figure}

\subsection{Geometry}

The geometry of the computational domain reflects an idealized
coronary artery. We show a sketch of the geometry in Figure~\ref{fig:domain}.  It
consists of three parts, two straight and one 
curved tube. The straight parts are aligned with the $x-$ and $y-$
axes for inflow and outflow, respectively.  
The centerline of the curved section is a part of a circle with center
in $(1,0,0)$ and radius $R=1$. It can be indicated as a parametrization
$\psi_C:[0,3]\to\mathds{R}^3$
\[
\psi_C(s)\coloneqq\begin{cases}
(0,-1+s,0)^T & 0\le s\le 1\\
(1+\cos(\pi s/2),\sin(\pi s/2),0)^T & 1\le s\le 2\\
(s-1,1,0)^T & 2\le s\le 3.
\end{cases}
\]
The fluid domain $\FL$ is a cylinder around the centerline $\psi_C(s)$
with radius $r_{\FL}=\unit[0.15]{cm}$. Furthermore, the solid domain (i.e. the vessel wall)
is a cylindrical outer layer of width $\unit[0.06]{cm}$,
i.e. the outer radius is given by $r_{\SO}=\unit[0.21]{cm}$. These
dimensions correspond the dimensions of realistic arteries.

We consider two stenotic and one non-stenotic configuration. The stenosis
is modelled  by a reduction of the inner radius $r_\FL$ on the second
part of the curved boundary. This can be described by the parametrization

\[
r_\FL^{sten}(s) \coloneqq \begin{cases}
  \unit[0.15]{cm} & 0\le s\le 1.5\\
  \unit[0.15]{cm}-\unit[0.02]{cm}\big(\cos(4\pi s)-1\big) & 1.5\le
  s\le 2\\ 
  \unit[0.15]{cm} & 2\le s\le 3.
\end{cases}
\]
Moreover we proceed twofold. First, we consider a symmetric stenosis
such that the fluid domain is a cylinder of radius $r_\FL(s)$ around
the centerline $\psi_C(s)$. Second, we modify the  centerline in such  a way 
that the stenosis is only on the inner side of the curve. This is achieved
by shifting the centerline to match the radius variation, i.e. 
\[
\psi_C^{shift}(s)\coloneqq\psi_C(s) -r_{\FL}^{sten}(s)\vec\chi(s),
\]
where $\chi\omega(s)$ is the direction of the shift
\[
\vec\omega(s) = \begin{cases}
(0,-0.02\big(\cos(4\pi s)-1\big),0) & 1.5\le s\le 2   \\
(0,0,0)^T &   \text{ elswhere }
\end{cases}.
\]
In both stenotic configurations, the resulting reduction of the radius
equals $\unit[0.04]{cm}$ and the lumen area shrinks by $46\%$ 
from $\unit[0.15^2\pi]{cm^2}\approx\unit[0.07]{cm^2}$ to
$\unit[0.11^2\pi]{cm^2}\approx \unit[0.038]{cm^2}$. 
Computational geometries are summarized in  
Table~\ref{tab:1}. Figure~\ref{fig:domain} shows Geometry 3. 

\begin{table}[t]
  \caption{Summary of geometry configurations. Figure~\ref{fig:domain}
    shows Geometry 3.} 
  \label{tab:1}       
  \begin{center}
  \begin{tabular}{l|cccc}
    \toprule
    &  description & centerline & inner radius & outer radius\\
    \midrule
    Geometry 1 & without stenosis
    & $\psi_C(s)$ 
    &$\unit[0.15]{cm}$
    &$\unit[0.21]{cm}$\\
    Geometry 2 & symmetric stenosis
    & $\psi_C(s)$ 
    &$r_{\FL}^{sten}(s)$
    &$\unit[0.21]{cm}$\\
    Geometry 3 & non-symmetric stenosis
    & $\psi_C^{shift}(s)$ 
    &$r_{\FL}^{sten}(s)$
    &$\unit[0.21]{cm}$\\
    \bottomrule
  \end{tabular}
  \end{center}
\end{table}

\subsection{Boundary conditions}

\begin{figure}[t]
    \begin{center}
        \begin{tikzpicture}
        \begin{axis}[
        title = {},
        width = 12cm,
        height = 5cm,
        axis x line=bottom, axis y line=left,
        enlarge x limits={abs=.1},
        enlarge y limits={abs=0.1},
        ylabel={$v_{\text{max}}(t)\;\unit{[cm/s]}$},
        tickpos=left,
        tick style = {semithick},
        grid=both,
        grid style={line width=.5pt, draw=gray!10},
        legend pos= south east,
        legend cell align={left},
        xtick={0, .1, .3, 1,  2, 3, 4},
        yticklabel style = {font=\tiny},
        xticklabel style = {font=\tiny},
        legend style={font=\small}
        ]

       
        \addplot[color= RoyalBlue, line width=.8pt, domain=0.1:0.3]
	plot (\x, { 28.3*3 } );
	\addplot[color= RoyalBlue, samples=100, smooth, line width=.8pt, domain=0.0:0.1]	
	plot (\x, {28.3* 1.5*(1.0-cos(deg(31.416*\x) ))  });		
	\addplot[color= RoyalBlue, samples=100, smooth, line width=.8pt, domain=0.3:3.3]	
	plot (\x, { 28.3*( 5.8269041571842815 + 0.106340253 -1.3933891269783132*cos(deg(2*3.14*1*(\x-0.3)))
	  +     1.3532413810149053*sin(deg(2*3.14*1*(\x-0.3))) -0.9409003055488833*cos(deg(2*3.14*2*(\x-0.3))) +
	  0.23325998681918006*sin(deg(2*3.14*2*(\x-0.3)))-0.3026878295161429*cos(deg(2*3.14*3*(\x-0.3))) +
	  -0.11904366029482762*sin(deg(2*3.14*3*(\x-0.3))) -0.22647020959258754*cos(deg(2*3.14*4*(\x-0.3))) +
	  -0.06313135613020682*sin(deg(2*3.14*4*(\x-0.3))) -0.10646173529097325*cos(deg(2*3.14*5*(\x-0.3))) +
	  -0.21375312564927748*sin(deg(2*3.14*5*(\x-0.3)))+ 0.040251489761758366*cos(deg(2*3.14*6*(\x-0.3))) +
	  -0.06918573256164762*sin(deg(2*3.14*6*(\x-0.3))) -0.030781624456619635*cos(deg(2*3.14*7*(\x-0.3))) +
	  -0.04510100577203846*sin(deg(2*3.14*7*(\x-0.3)))+ 0.027194931202571635*cos(deg(2*3.14*8*(\x-0.3))) +
	  -0.07355363347737506*sin(deg(2*3.14*8*(\x-0.3))) )   });
	\addlegendentry{ maximal velocity}
        \end{axis}
        \end{tikzpicture}
    \end{center}
    
    \caption{Inflow profile. The exact flow rate is given
      in~\eqref{inflow}.}
    \label{fig:inflow}
\end{figure}
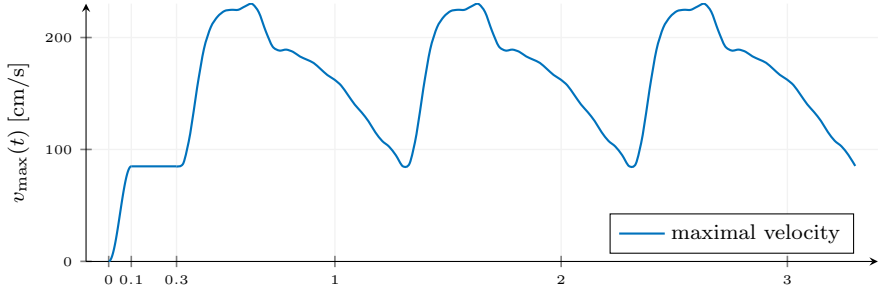

The blood flow is enforced by a Dirichlet inflow condition on the inflow
boundary $\Gamma_f^{in}$ given by
\[
\Gamma_f^{in}=\{ (x,-1,z)\in\mathds{R}^3\,|\,
\sqrt{x^2+z^2}<\unit[0.15]{cm}\}. 
\]
We prescribe a parabolic inflow profile
\[
\vt^{in}(x,y,z,t)=v_\text{max}(t)  \Big(
1-\frac{x^2}{\unit[0.15^2]{cm^2}}-\frac{z^2}{\unit[0.15^2]{cm^2}}\Big)
(0,1,0)^T,
\]
where the maximum value $v_{\text{max}}(t)$ is presented in 
Figure~\ref{fig:inflow}. The temporal inflow profile 

\begin{equation}\label{inflow}
v_\text{max}(t) = \frac{1}{28.3}\begin{cases}
0.5(1.0-\cos(\pi t/0.1))3.0 & 0\le t\le 0.1\\
3.0 & 0.1\le t \le 0.3\\
\begin{split}
&5.931\\
&-1.3933\cos(2\pi 1t)+1.3532\sin(2\pi 1t) \\
&-0.9409\cos(2\pi 2t)+0.2332\sin(2\pi 2t) \\
&-0.3026\cos(2\pi 3t)-0.1190\sin(2\pi 3t) \\ 
&-0.2264\cos(2\pi 4t)-0.0631\sin(2\pi 4t) \\
&-0.1064\cos(2\pi 5t)-0.2137\sin(2\pi 5t) \\
&+0.0402\cos(2\pi 6t)-0.0691\sin(2\pi 6t) \\
&-0.0307\cos(2\pi 7t)-0.0451\sin(2\pi 7t) \\
&+0.0271\cos(2\pi 8t)-0.0735\sin(2\pi 8t) 
\end{split}& 0.3\le t\le 3.3
\end{cases}
\end{equation}
is an 
approximation, by means of a~Fourier series of order 8, of a typical coronary velocity profile provided by~\cite{hemolab}. The inflow profile consists of three stages:
\begin{itemize}
    \item Ramp phase $(\unit[0]{s}\le t\le\unit[0.1]{s})$ with
      increasing inflow rate.
    \item Steady phase $(\unit[0.1]{s}<t\le\unit[0.3]{s})$ with
      constant inflow rate. 
    \item Pulsatile phase $(\unit[0.3]{s} <t)$ with pulsatile inflow
      corresponding to heartbeats.
\end{itemize}

For the outflow of the fluid on $\Gamma_f^{out}$ in the FSI case we
apply absorbing boundary conditions:

\begin{equation}\label{abs_neuman}
(\sigma_f^{n+1}\cdot \n)|_{\Gamma_{out}} =  
\left(\left( \frac{\sqrt{\rho_f}}{2\sqrt{2}}\frac{Q^n}{A^n} 
+ \sqrt{p^*} \right)^2 + p_e - p^*\right)\cdot \n ,
\end{equation}
where $p* = \frac{E}{1-\nu^2}$. 
The condition is equipped with the reference pressure $p_e$ which is chosen such that a vascular pressure of $\unit[80]{mmHg}$ at $t=\unit[0.3]{s}$ is achieved. The absorbing condition explicitly disallows
pressure waves to reenter the domain, for details see~\cite{Nobile2008}.

The structure is fixed at the inflow boundary and is allowed to move in $y$-$z$ direction at the outflow boundary.

In order to compare both rigid and elastic computations, we run the FSI simulation first and then prescribe the resulting reference pressure profile $P_{ref}(t)$ at the outflow boundary in the rigid wall case such that 
\[
\big\langle\big(
\rho_f\nu_f\nabla\vt_f-pI\big)\nt,\phit\big\rangle_{\Gamma_f^{out}}
=
\big\langle
P_{ref}(t)\nt,\phit\big\rangle_{\Gamma_f^{out}}. 
\]
This is the usual do-nothing outflow condition including a pressure
offset, see~\cite{HeywoodRannacherTurek1992}.

The first two phases of \eqref{inflow} are designed to reach the intravascular 
pressure of $\unit[80]{mmHg}$. This procedure resembles prestressing 
of the vessel wall. We first start the FSI simulation on the undeformed stress free reference domain and prestress 
the geometry by the ramp and steady phases. The resulting geometry at $t=\unit{0.3}{s}$ is then extracted to 
compute the FSI as well as the Navier-Stokes simulation on this deformed geometry. This geometry 
corresponds to a reconstructed artery geometry from MRI or CT images, as during the measurements the blood flows 
with the pressure of at least $\unit[80]{mmHg}$ through the blood vessel.
As the stress free reference domain is known here a 
priori, prestressing procedures 
as discussed in \cite{gee2010computational} do not need to be applied.


\section{Numerical approximation, solution and
  implementation}\label{sec:nummethod}

The realization of a numerical framework for monolithic fluid-structure
interactions is very challenging and a detailed description is not
possible in one manuscript. We therefore give a brief description and
refer to the relevant and detailed literature. See~\cite{Richter2017}
for a comprehensive overview.

We employ a very strict form of the ALE formulation, where all
equations are solved on the reference domain, no mesh update is
used. This prevents the necessity of projections between moving meshes
and also it is the only straightforward approach for obtaining higher
order discretization in time~\cite{RichterWick2015_time}. In principle
this approach allows for direct Galerkin discretizations of the
monolithic variational formulation~(\ref{probelm}) and the choice of
finite element spaces should be based on the following considerations
\begin{itemize}
\item The finite element mesh should resolve the fluid-structure
  interface $\IN$ in reference framework. 
\item In the fluid domain, the velocity-pressure pair $V_h\times Q_h$
  should fulfils the inf-sup condition to cope with the
  incompressibility constraint. Or, if a non-stable finite element
  pair is used, stabilization terms must be added. Our realization is
  based on equal order tri-quadratic elements for pressure and
  velocity, enriched with the local projection stabilization
  method~\cite{BeckerBraack2001}. 
\item To reach a balanced approximation of the velocity-deformation pairing
  the same function space is used for the global deformation field. 
\end{itemize}
To summon up, the discrete solution $U_h\coloneqq(\vt_h,\ut_h,p_h)$ is found
in the space $X_h=[V_h]^3\times [V_h]^3\times V_h^f$, where $V_h$
extends over the complete domain and $V_h^f$ over the fluid domain
only.

In time we use a variant of the Crank-Nicolson time discretization
scheme that gives better stability properties by an implicit
shifting. We refer to~\cite{RichterWick2015_time} for details.  
By restricting~(\ref{probelm}) to the fully discrete setting, a system
on nonlinear algebraic equations arises in each time step. As
nonlinear solver we employ a 
Newton scheme with an analytical evaluation of the Jacobian,
see~\cite{Richter2012a} or~\cite[Section 5.2.2]{Richter2017} for
details on the derivation. 

The resulting linear systems of equations are very large and extremely
ill-conditioned with condition numbers that are by far larger than
those of fluid and solid equation on their own,
see~\cite{Richter2015,AulisaBnaBornia2018} for numerical studies. The
approximation of these systems is still  a great challenge, in
particular if it comes to 3d 
applications. Only few fast solvers for the nonlinear setting are
available~\cite{Richter2015,AulisaBnaBornia2018,JodlbauerLanger}.
Our approach is based on a multigrid solution that appears to be
superior in 3d. In~\cite{FailerRichter2020} we present the solution
approach, which is based on a partitioning of the Jacobain based on
two simple strategies:
\begin{enumerate}
\item Within the Navier-Stokes equations we neglect those parts of the
  Jacobian that come from the derivative with respect to the domain
  extension $\ut_f$. The resulting nonlinear solver is an approximated
  Newton method that however still solves the original problem since
  the residual is exact. In~\cite[Section 5.2.3]{Richter2017} and
  \cite{FailerRichter2020} we have found that
  such an approximated Newton solver is even more efficient,
  despite slightly increased iteration counts. This is due to the very
  costly evaluation of the full Jacobian that is not required in our
  approach.   
\item We exploit the discretization of the equation $d_t\ut_s=\vt_s$,
  precisely, if we consider the Crank-Nicolson scheme
  \[
  \ut^n_s=\ut^{n-1}_s + \frac{k}{2}
  \Big(
  \vt_s^{n-1}
  +\vt_s^{n}\Big) \eqqcolon {\cal U}(\vt_s^n;\ut^{n-1}_s,\vt_s^{n-1}),
  \]
  and we replace the dependency of the solid stress tensor
  on the deformation by  the velocity, i.e.
  \[
  \Sigmat_s(\ut_s) = \Sigmat_s\big({\cal
    U}(\vt_s^n;\ut^{n-1}_s,\vt^{n-1}_s)\big). 
  \]
\end{enumerate}
The combination of these two modifications allows for a natural splitting
of the Jacobian within the multigrid smoother. Further, it allows to
apply very simple iterations of Vanka-type, as smoother in the geometric multigrid preconditioner, that are easy to
parallelize. In~\cite{FailerRichter2020} we have demonstrated the
efficiency of the approach in different 3d configurations.

The implementation is based on the finite element toolkit 
Gascoigne3D~\cite{Gascoigne3D}. 

\section{Simulations}\label{sec:sim}

We finally use the presented finite element framework for a
computational analysis of several hemodynamic parameters that are
relevant to answer clinical questions. We focus on the dependence of these parameters on the elasticity of the vessel walls. The additional effort of fully coupled fluid-structure interactions over pure
Navier-Stokes simulations is immense and should be justified by
corresponding effects.

The behavior of three specific parameters is investigated. The wall shear stress (WSS)
plays an eminent role in several applications. It is used as indicator
to model the growth of atherosclerotic
plaques~\cite{YangJaegerNeussRaduRichter2015,RichterMizerski2020}
but also when it comes to deriving measures to evaluate the risk of plaque
rupture~\cite{Libby2019} or the rupture of aneurysms~\cite{Cebral2010}.
Both the minimum wall shear stress and the distribution of 
the wall shear stress on the vessel walls are important measures. 
In Section~\ref{num:wss} analyze the effect of the different 
complexities, Navier-Stokes and fluid-structure interactions, 
on the WSS distribution in all three different geometries. 

Furthermore we analyze the \emph{computational fractional
flow reserve} (cFFR). The \emph{fractional flow reserve} (FFR) is a
technique that measures pressure differences across a stenosis. In
medical practice, the FFR is determined by inserting a catheter in the
artery and measuring flow parameters at maximum blood flow. During the
procedure, the tip of the catheter, where the sensor is
located, is retrieved, such that measurements are available along the
affected section of the vessel. Healthy vessels should give a pressure
ratio close to 1. If the ratio drops below $0.8$, i.e. a $20\%$ drop
in pressure, the stenosis is considered to be severe~\cite{Tonino2009}. 
Aim of the \emph{computational fractional flow reserve} is to replace
the risky intervention by computer simulations based on medical
imaging. 

Finally, 
we discuss the amplitude of the
pressure oscillation during one heart cycle. In clinical observation
is usually observed that the amplitude significantly drops after a
stenotic region in a blood vessel. By comparing Navier-Stokes
simulations with coupled fluid-structure interactions, we will show
that this effect can only be described by considering the fully
coupled model including elastic vessel walls.

\subsection{Wall shear stress (WSS)}\label{num:wss}

\begin{figure}[t]
  \centerline{\includegraphics[width=1.\textwidth]{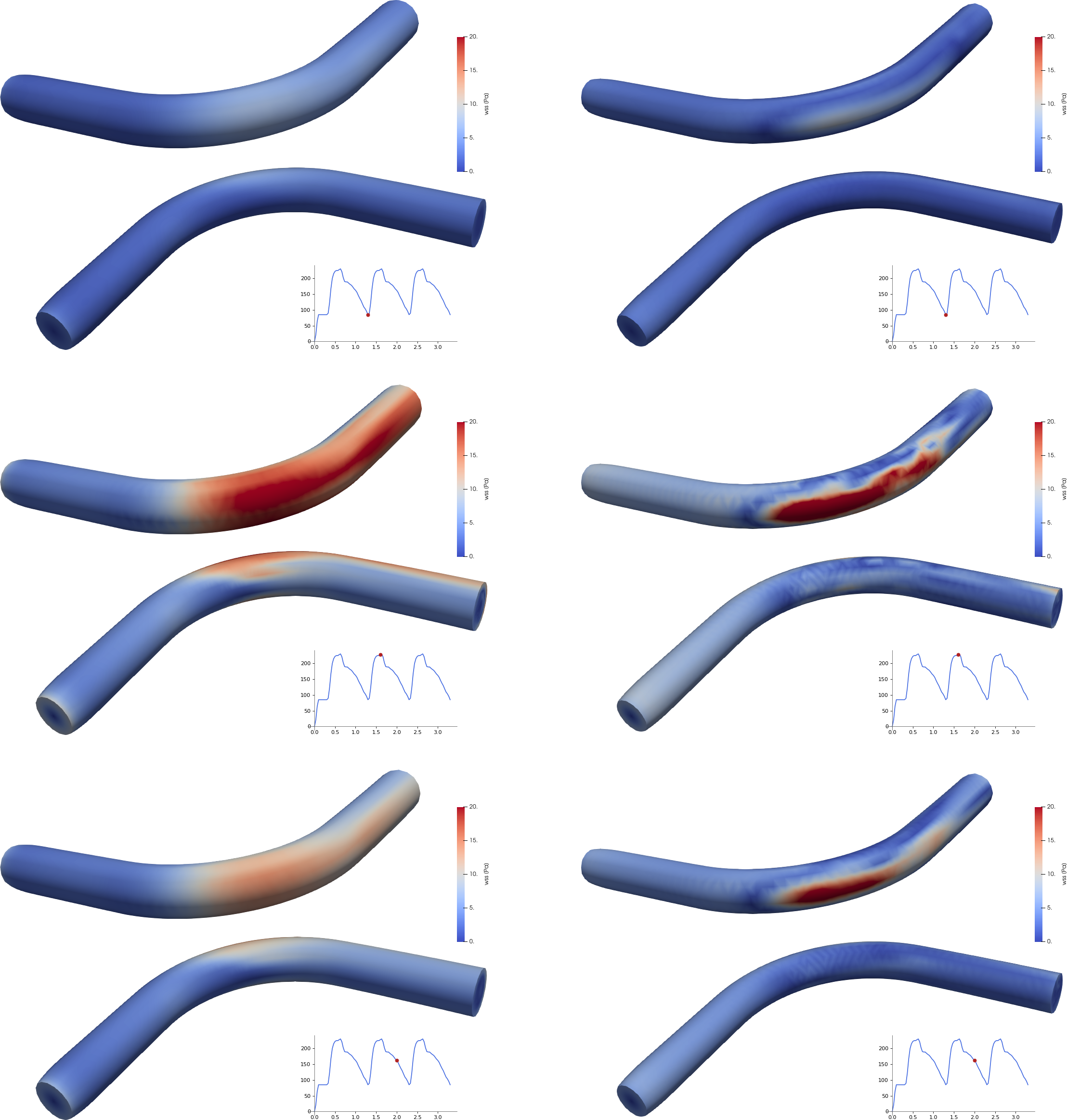} }
  \caption{Wall Shear stress for Geometry 1 (without  stenosis). On
    the left FSI right NS.
    We show each configuration form two different perspectives, such
    that inner and outer region of the curved arch are visible. 
  }
  \label{fig:wsscolonge}       
\end{figure}

The tangential component of the surface force at the vessel
wall is denoted as wall shear stress (WSS). By means of the Cauchy's theorem we have
\begin{equation}\label{wss}
  \wss =  (\sigmat_f(\vt,p) \nt\cdot \taut)  \taut =[I-\nt\nt^T]\sigmat_f(\vt,p) \nt ,
\end{equation}
where $\n$ is a unit outward normal vector to  the vessel wall and
$\taut$ is corresponding tangential vector.  Note that WSS is a vector
and its often confused with its magnitude $|\wss|_2$ which is a scalar  
quantity denoted by $wss$.

\begin{figure}[t]
  \centerline{\includegraphics[width=1.\textwidth]{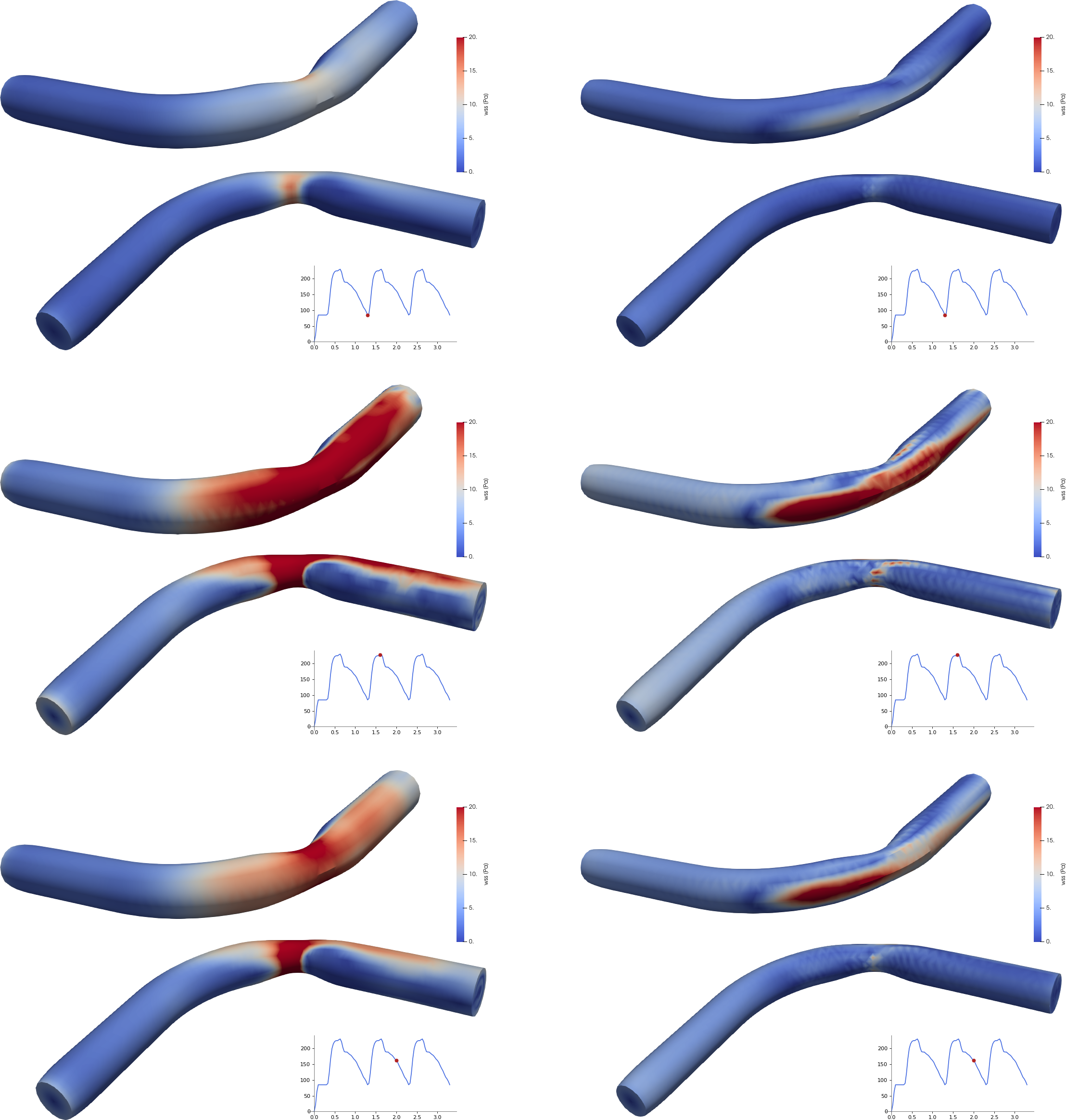} }
  \caption{Wall Shear stress for Geometry 2 (non-symmetric stenosis). On
    the left FSI right NS.
    We show each configuration form two different perspectives, such
    that inner and outer region of the curved arch are visible. }
  \label{fig:wssnonsymsten}       
\end{figure}

\begin{figure}[t]
  \centerline{\includegraphics[width=1.\textwidth]{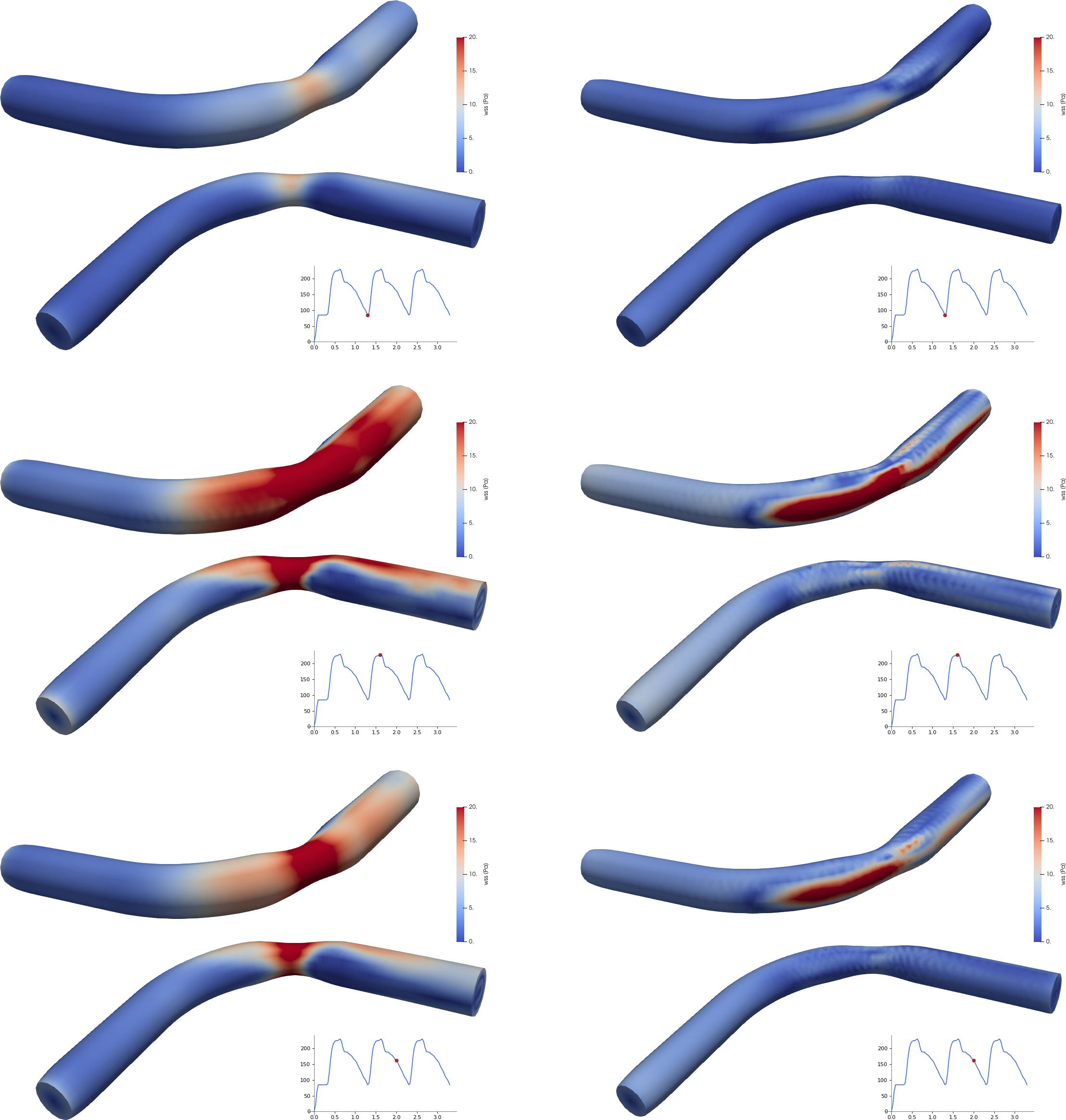} }
  \caption{Wall Shear stress for Geometry 3 (symmetric stenosis). On
    the left FSI right NS.
    We show each configuration form two different perspectives, such
    that inner and outer region of the curved arch are visible. }
  \label{fig:wsssymsten}       
\end{figure}

The plots of the wall shear stress are presented on the boundary of
the fluid domain, see Figure~\ref{fig:wsscolonge},
Figure~\ref{fig:wsssymsten} and Figure~\ref{fig:wssnonsymsten}. The
surrounding solid is removed from these plots such that only the fluid
domain is given.

We use the same scale ranging from $\unit[0]{Pa}$ to $\unit[20]{Pa}$
in all figures. The distribution of $wss$ is always presented for
3 different points in time. First, when the pulsatile flow reaches its
minimum inflow pressure, then, at maximum pressure and finally for an
intermediate pressure value. Each specific situation is shown from two
different angles, such that the inner and outer surfaces are well
visible. 

Comparing the coupled fluid-structure interaction model (left) with
the pure Navier-Stokes flow (right) always shows much higher
wss values in the Navier-Stokes case. Regions of very high
$wss$ are only found in the outer surface in the case of
Navier-Stokes. In the case of the FSI model, the values are smoothly
spread. In general, the vessels are widened and show a larger diameter
of the lumen in the case of FSI. Prestressing of the domains for the
Navier-Stokes simulations is based on the deformation resulting from
the increasing inflow of the ramp phase~\eqref{inflow} to reach 
pressure of $80\unit{mmHg}$.

Figures~\ref{fig:wsssymsten} and~\ref{fig:wssnonsymsten} show a  shift
of the wall shear stress distribution. High values are found all
around the stenosed parts, on the inner and the outer surfaces of the
curved areas. Again, the Navier-Stokes model is not able to yield a proper
distribution of $wss$ around the cylinder surface and it
only concentrates in the outer surface of the curved section.

To sum up, it appears to be important to consider elastic
fluid-structure interactions, if the spatial distribution of the wall
shear stress is of interest. The Navier-Stokes model is nearly unaware
of the stenosis and always concentrates the WSS on the outer surface
of the curved region. If rupture locations~\cite{Cebral2010} or localized growth
processes are of
interest~\cite{YangJaegerNeussRaduRichter2015,RichterMizerski2020},
the use of FSI is essential.


\subsection{Computational fractional flow reserve (cFFR)}
\label{num:ffr}

The \emph{computational fractional flow reserve} (cFFR) is the ratio
of the maximum blood pressure after the stenosis (distal) and before
the stenosis (proximal). We denote by $p_d$ the distal pressure and by
$p_a$ the proximal pressure such that the computational fractional
flow reserve is defined by
\[
\text{cFFR} = \frac{p_d}{p_a}. 
\]
In our configuration, compare with Figure~\ref{fig:domain}, we evaluate the
distal and proximal values in
\begin{equation}\label{pcord}
  p_a=p(0,0,0)\text{ and } p_d=p(1.7,1,0).
\end{equation}

Since the cFFR varies slightly with time the maximum value over the cardiac cycle
is computed. Further computational aspects
of cFFR are covered by the recent benchmarking paper~\cite{Carson2019}. 
In healthy vessels we expect $\text{cFFR}\approx 1$ and in the medical
practice, a stenosis with 
$\text{cFFR} > 0.8$ is considered to be functionally
non significant~\cite{Tonino2009}. No medical intervention, e.g. by placing a
stent would be required. 

The results of cFFR computations are presented in
Table~\ref{tab:cffr}. The first striking observation is that the
Navier-Stokes model is not able to reflect the presence of the
stenosis at all. A pressure drop of about $5\%$ is observed for all
configurations and is only due to the curvature of the domain. The FSI
model is well able to yield $\text{cFFR}\approx 1$ in the case of the
healthy configuration and shows a loss of about $5\%$ in both stenotic
geometries. Based on these simulations, the stenosis would not be
considered to be severe and in the need of an intervention. These results
clearly show that a pure Navier-Stokes simulation is not able to serve
as computational basis for replacing the medical FFR procedure by
simulations. 

\begin{table}[t]
  \caption{Results of the cFFR analysis in all three geometries,
    considering the fully coupled fsi model and the pure Navier-Stokes
    case. }
  \label{tab:cffr}       
  \begin{center}
    \begin{tabular}{lccc}
      \toprule
      &  description & \textit{FSI} & \textit{NS} \\
      \midrule
      Geometry 1 & without stenosis       & 0.99 & 0.96\\
      Geometry 2 & non-symmetric stenosis & 0.96 & 0.94\\
      Geometry 3 & symmetric stenosis     & 0.96 & 0.95\\
      \bottomrule
    \end{tabular}
  \end{center}
\end{table}

\subsection{Pressure amplitude}\label{num:pressure}
\begin{figure}[t]
    \begin{center}
        \begin{tikzpicture}
        \begin{axis}[
        title = {FSI},
        width = 6cm,
        height = 5cm,
        axis x line=bottom, axis y line=left,
        enlarge x limits={abs=.05},
        ylabel={ without stenosis},
        tickpos=left,
        tick style = {semithick},
        grid=both,
        grid style={line width=.5pt, draw=gray!10},
        legend pos= outer north east,
        legend cell align={left},
        ymin=80,ymax=124,
        yticklabel style = {font=\tiny},
        xticklabel style = {font=\tiny},
        legend style={font=\tiny}
        ]
        \addplot[color= OliveGreen, line width=.8pt 
        ] table [x=time, y=pacfsi, col sep=tab] {./results/pressure.dat};
        \addplot[color= BurntOrange, line width=.8pt
        ] table [x=time, y=pdcfsi, col sep=tab] {./results/pressure.dat};
        \end{axis}
        
        \tikzset{shift={(6,0)}}
        
         \begin{axis}[
        title = {NS},
        width = 6cm,
        height = 5cm,
        axis x line=bottom, axis y line=left,
        enlarge x limits={abs=.05},
        ylabel={pressure [mmHg]},
        tickpos=left,
        tick style = {semithick},
        grid=both,
        grid style={line width=.5pt, draw=gray!10},
        legend pos= outer north east,
        legend cell align={left},
        ymin=80,ymax=124,
        yticklabel style = {font=\tiny},
        xticklabel style = {font=\tiny},
        legend style={font=\tiny}
        ]
        \addplot[color= OliveGreen, line width=.8pt 
        ] table [x=time, y=pacns, col sep=tab] {./results/pressure.dat};
        \addplot[color= BurntOrange, line width=.8pt
        ] table [x=time, y=pdcns, col sep=tab] {./results/pressure.dat};
        \end{axis}
        
        \tikzset{shift={(-6,-4.)}}
        
        \begin{axis}[
        title = {},
        width = 6cm,
        height = 5cm,
        axis x line=bottom, axis y line=left,
        enlarge x limits={abs=.05},
        ylabel={non-symmetric stenosis},
        tickpos=left,
        tick style = {semithick},
        grid=both,
        grid style={line width=.5pt, draw=gray!10},
        legend pos= outer north east,
        legend cell align={left},
        ymin=80,ymax=124,
        yticklabel style = {font=\tiny},
        xticklabel style = {font=\tiny},
        legend style={font=\tiny}
        ]
        \addplot[color= OliveGreen, line width=.8pt 
        ] table [x=time, y=panssfsi, col sep=tab] {./results/pressure.dat};
        \addplot[color= BurntOrange, line width=.8pt
        ] table [x=time, y=pdnssfsi, col sep=tab] {./results/pressure.dat};
        \end{axis}
        
        \tikzset{shift={(6,0)}}
        
        \begin{axis}[
        title = {},
        width = 6cm,
        height = 5cm,
        axis x line=bottom, axis y line=left,
        enlarge x limits={abs=.05},
        ylabel={pressure [mmHg]},
        tickpos=left,
        tick style = {semithick},
        grid=both,
        grid style={line width=.5pt, draw=gray!10},
        legend pos= outer north east,
        legend cell align={left},
        ymin=80,ymax=124,
        yticklabel style = {font=\tiny},
        xticklabel style = {font=\tiny},
        legend style={font=\tiny}
        ]
        \addplot[color= OliveGreen, line width=.8pt 
        ] table [x=time, y=panssns, col sep=tab] {./results/pressure.dat};
        \addplot[color= BurntOrange, line width=.8pt
        ] table [x=time, y=pdnssns, col sep=tab] {./results/pressure.dat};
        \end{axis}
        \tikzset{shift={(-6,-4)}}
        
        \begin{axis}[
        title = {},
        width = 6cm,
        height = 5cm,
        axis x line=bottom, axis y line=left,
        enlarge x limits={abs=.05},
        ylabel={symmetric stenosis},
        tickpos=left,
        tick style = {semithick},
        grid=both,
        grid style={line width=.5pt, draw=gray!10},
        legend pos= outer north east,
        legend cell align={left},
        ymin=80,ymax=124,
        yticklabel style = {font=\tiny},
        xticklabel style = {font=\tiny},
        legend style={font=\tiny}
        ]
        \addplot[color= OliveGreen, line width=.8pt 
        ] table [x=time, y=passfsi, col sep=tab] {./results/pressure.dat};
        \addplot[color= BurntOrange, line width=.8pt
        ] table [x=time, y=pdssfsi, col sep=tab] {./results/pressure.dat};
        \end{axis}
        
        \tikzset{shift={(6,0)}}
        
        \begin{axis}[
        title = {},
        width = 6cm,
        height = 5cm,
        axis x line=bottom, axis y line=left,
        enlarge x limits={abs=.05},
        ylabel={pressure [mmHg]},
        tickpos=left,
        tick style = {semithick},
        grid=both,
        grid style={line width=.5pt, draw=gray!10},
        legend style={at={(-.2,-.3)},anchor=south},
        legend columns = 2,
        ymin=80,ymax=124,
        yticklabel style = {font=\tiny},
        xticklabel style = {font=\tiny},
        ]
        \addplot[color= OliveGreen, line width=.8pt 
        ] table [x=time, y=passns, col sep=tab] {./results/pressure.dat};
        \addlegendentry{ $p_a$}
        \addplot[color= BurntOrange, line width=.8pt
        ] table [x=time, y=pdssns, col sep=tab] {./results/pressure.dat};
        \addlegendentry{ $p_d$}
        \end{axis}
        \end{tikzpicture}
    \end{center}
    
    \caption{Pressure waves for one cardiac cycle }
    \label{fig:pressure}
    
\end{figure}
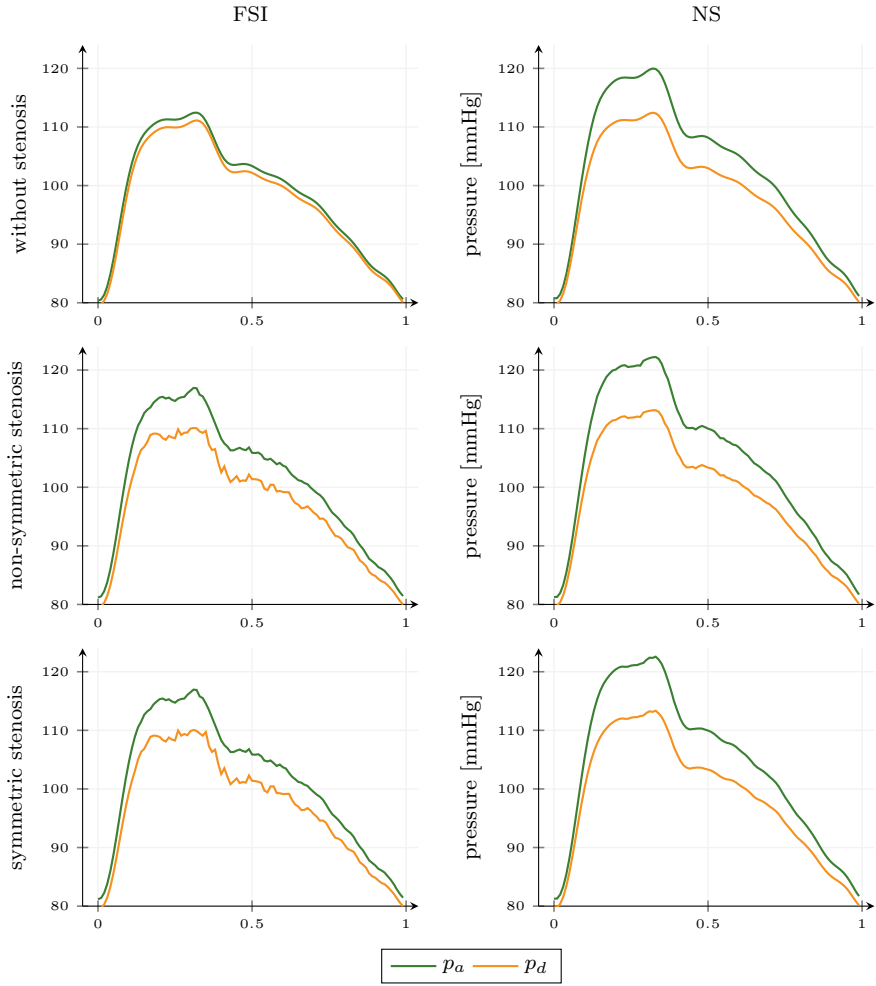

The third quantity of interest is the dynamic pressure amplitude
before and after the stenosis, i.e. a temporally resolved analysis of
the proximal and distal pressures $p_a$ and $p_d$ evaluated in the
coordinates as mentioned in~\eqref{pcord}. We study the progress of
the pressure to investigate possible damping effects of a stenosis. 
The change of pressure profile after strongly stenotic regions, 
is reported as an assessment tool for cardiovascular risk \cite{Pichler2015}.

We show the evolution of the pressures $p_a(t)$ and $p_d(t)$ over the
third heart beat in Figure~\ref{fig:pressure}. On the left, we show
results for the fully coupled FSI models, on the right we give the
corresponding pressure lines for the Navier-Stokes case. 
Comparable to the cFFR
study given in the previous section, the most striking observation is
the invariance of the Navier-Stokes solution to the kind of vessel
geometry. For all cases, healthy vessel, centered stenosis and non
symmetric stenosis, the Navier-Stokes results are nearly identical and
always indicate a loss in pressure amplitude of about $7.5\%$. In
contrast, the FSI solution is able to preserve the oscillation in the
case of the healthy vessel and shows a drop of about $5\%$ for the two
stenotic cases. A closer look reveals that the distal pressure lines
are very similar in all cases. The slight oscillations in the distal
pressure in case of the FSI simulation are not numerical
instabilities. Instead they stem from the oscillatory behavior of the
elastic vessel wall. 

Finally, Table~\ref{tab:pressure} collects the amplitudes before and
after the stenosis. 
This third study also shows enormous qualitative and quantitative
discrepancies between the simulation results depending on the model
under consideration. 

\begin{table}[t]
    \caption{Pressure Amplitude in $\unit{mmHg}$. We compare the drop
      of the amplitude over the stenotic region and compare the
      results for pure Navier-Stokes flow (left) with fully coupled
      fluid-structure interactions (right).  }
    \label{tab:pressure} 
    \begin{center}
      \begin{tabular}{lccc|ccc}
        \toprule
        &\multicolumn{3}{c|}{Navier-Stokes}
        &\multicolumn{3}{c}{FSI}\\
        &  proximal& distal& drop& proximal&distal&drop\\
        \midrule
        Geometry 1 & 39.2 & 32.7& 7.5 & 32.1 & 31.4&0.7 \\
        Geometry 2 & 41.0 & 33.4& 7.6 & 35.8 & 30.5&5.3 \\
        Geometry 3 & 41.3 & 33.6& 7.7 & 35.8 & 30.4&5.4 \\
        \bottomrule
      \end{tabular}
    \end{center}
\end{table}

\section{Conclusions}\label{sec:con}

We have studied a prototypical geometry that represents a large and
curved artery. Three different configurations are considered: a
healthy vessel with a diameter (the lumen) of $\unit[0.15]{cm}$ and
two cases where a stenosis is given in the curved area of the
vessel. First, the stenosis is centered around the vessel centerline,
finally, a shifted variant where the stenosis is concentrated on the
inner surface. The blood flow is driven by a time-dependent flow rate
with values that are clinically relevant. For all configurations we
study the effect of the elasticity in the vessel walls, i.e. we
compare a pure Navier-Stokes simulation with a fully coupled
fluid-structure interaction system.

Three different indicators that are also relevant in clinical decision
making are investigated: the distribution of the wall shear stress
that is made responsible for stenosis growth and risk of rupture, the
computational fractional flow reserve that is used to estimate the
severity of a stenosis and the amplitude of the pressure oscillation
that also measures the severity of a stenosis. In all cases we
observe that the simple Navier-Stokes model is not able to depict the
effect of the plaque. In particular the pressure lines are nearly
identical for all three geometrical configurations. The FSI model is
however well able to replicate clinical observations. For instance,
the energy, in terms of pressure oscillations, is fully preserved
throughout the curved region, if elasticity of the vessel walls is
taken into account.

Although the study covers only an idealised geometry and boundary conditions,
the proposed  regime corresponds to physiological blood flow. Therefore,
we stress that the compliance of the vessel has significant impact 
on clinical hemodynamical factors. In medical practise one has to be
aware of the difference in the results from CFD  with FSI and NS
models and the limitations of the later.

\begin{acknowledgements}
The authors acknowledge the financial support by the Federal Ministry of
Education and Research of Germany, grant number 05M16NMA. PM and TR 
acknowledge the support of the GRK 2297 MathCoRe, funded by the Deutsche
Forschungsgemeinschaft, grant number 314838170. 
\end{acknowledgements}

%
\section*{Conflict of interest}
The authors declare that they have no conflict of interest.


\end{document}